
\magnification=1200  
\def\newline{\hfill\penalty -10000}  
\def\title #1{\centerline{\rm #1}}
\def\author #1; #2;{\line{} \centerline{#1}\smallskip\centerline{#2}}
\def\abstract #1{\line{} \centerline{ABSTRACT} \line{} #1}
\def\heading #1{\line{}\smallskip \goodbreak \centerline{#1} \line{}}
\newcount\refno \refno=1
\def\refjour #1#2#3#4#5{\noindent    \hangindent=1pc \hangafter=1
 \the\refno.~#1, #2 ${\bf #3}$, #4 (#5). \global\advance\refno by 1\par}
\def\refbookp #1#2#3#4#5{\noindent \hangindent=1pc \hangafter=1
 \the\refno.~#1, #2 (#3, #4), p.~#5.    \global\advance\refno by 1\par}
\def\refbook #1#2#3#4{\noindent      \hangindent=1pc \hangafter=1
 \the\refno.~#1, #2 (#3, #4).           \global\advance\refno by 1\par}
\newcount\equatno \equatno=1
\def\adveqn{(\the\equatno) \global\advance\equatno by 1}

\def\up#1{\leavevmode \raise 0.2ex\hbox{#1}}

\newcount\figno
\figno=0
\def\figure{\global\advance\figno by 1 Figure~\the\figno.~}
%

%
\vsize=8.75truein
\hsize=5.75truein
\hoffset=0.5truein
%
\baselineskip=0.166666truein
%
\parindent=25pt
%
\parskip=0pt
%
\nopagenumbers

%
%
\line{}

\title{
HYDRODYNAMICS AND HIGH-ENERGY PHYSICS OF}
\title{WR COLLIDING WINDS}

\author
Vladimir V. Usov \parindent=0pt ;
Dept.~of Physics, Weizmann Institute, Rehovot 76100, Israel ;

\abstract{The stellar winds flowing out of the components
of WR + OB binaries can collide and the
shock waves are formed. Stellar wind collision, particle acceleration
by the shocks and generation of X-ray, $\gamma$-ray, radio and IR
emission in WR + OB binaries are discussed.}

\heading{1. INTRODUCTION}
One of the main properties of Wolf-Rayet (WR) stars is a very
intense outflow of gas. The mass-loss rate for WR stars,
$\dot M_{_{WR}}$, and the terminal velocity of the matter outflow,
$V^{^\infty}_{_{WR}}$, far from the star amount to $\sim(0.8-8)\times
10^{-5}$ $M_\odot$ yr$^{-1}$ and $\sim(1-5)$$\times$10$^3$ km s$^{-1}$,
respectively (Willis 1982; Abbott et al. 1986; Torres, Conti and Massey
1986).  No less than 40\% \ of WR stars belong to binary systems.  Young
massive O and B stars are the secondary components of such systems.
OB stars also have an intense stellar wind: $\dot M_{_{OB}}\sim 10^{-6}
\,M_\odot$ yr$^{-1}, \,
V^{^\infty}_{_{OB}}\sim (1-3)\times 10^3$ km s$^{-1}$
(Garmany and Conti 1984; Leitherer 1988).
\par
If the intensities of the stellar winds of WR and OB stars are
more or less comparable or if the
distance $D$ between the components of the binary is large enough
(see below), the winds flowing out of WR and OB stars can collide
and the shock waves are formed. In the shock the gas is heated to
temperature $\sim 10^7$ K and generates X-ray emission (Prilutskii
and Usov 1975, 1976; Cherepashchuk 1976; Cook, Fabian and Pringle
1979; Bianchi 1982; Luo, McCray and Mac Low 1990; Usov 1990;
Stevens, Blondin and Pollock 1992; Usov 1992; Myasnikov and Zhekov 1993)
\par
Until now X-ray emission from a few tens WR stars has been detected
(Seward et al. 1979; Moffat et al. 1982; Caillaut et al. 1985; Pollock
1987). From the observational data it follows that the WR binaries are
significantly brighter in X-rays than  single stars (Pollock 1987), and
the WR stars detected in X-rays have a much higher than usual
incidence of binaries (Abbott and Conti 1987). These two facts can be
naturally explained if the X-ray emission of WR binaries is enhanced
essentially by the radiation of the gas heated in the shock.
Besides, the analysis of the energy
spectra of the X-ray emission and the time variability indicates that the
X-rays are generated in the outer regions of the stellar wind near the
OB star (Moffat et al. 1982; Pollock 1987; Williams et al. 1990), as
expected in the model of stellar wind collision (see Section 2).
\par Stellar wind collision may be responsible not only for the X-ray
emission of WR + OB binaries and for their radio, IR and $\gamma$-ray
emision as well (Williams, van der Hurcht and Th\' e 1987;
Williams et al. 1990, 1992, 1994; Usov 1991; Eichler and Usov 1993).
Below, the gas flow, stellar wind collision, particle acceleration, and
generation of X-ray, $\gamma$-ray, radio and IR emission in WR + OB
binaries are discussed.

\heading{2. STELLAR WIND COLLISION IN WR + OB BINARY}
{2.1 {\it Stellar wind parameters}
\smallskip
\par For a typical WR + OB binary $(\dot M_{_{WR}}\simeq 2 \times
10^{-5}M_\odot{\rm
yr}^{-1},\,\,\dot M_{_{OB}}\,\simeq\,10^{-6}M_\odot {\rm yr}^{-1},\,\,
V^{^\infty}_{_{WR}}\simeq V^{^\infty}_{_{OB}}\simeq 2\times 10^3\,{\rm
km}\,{\rm s}^{-1}$, the distance between the components of the binary
$D\simeq 10^{13}$ cm, the stellar wind gas
temperature  $T\,$$\simeq 3\times
$$10^4$ K) the parameters of the gas ahead of the shock fronts are
the following: The gas density $\rho \,$$\simeq$$ \,
10^{-14} \,{\rm g\; cm}^{-3}$,
the sound speed $V_s \,$$\simeq$$\,
10$ km s$^{-1}$, the free path length of charged
particles $l\,$$\simeq$$ \, 10^2$ cm, the Mach number $\xi =
V^{^\infty}_{_{WR}}/V_s\simeq 10^2\gg 1$, and the Reynolds number
$\Re =(r_{_{OB}}/l)\,m \simeq 10^{12}$. The free  path length
of particles and the Reynolds number behind the shoch fronts  are
$\sim 10^9$ cm and $\sim 10^3$, respectively. The above values of $\Re$
suggest that the gas flow  is  non-viscous and non-heat conductive.
\medskip
{2.2. {\it Geometry of the region of stellar wind collision}}
\smallskip
\par The winds from the binary components are highly supersonic
and flow nearly radially out to
the shocks. In the shock the gas is heated to the temperature
$$T(V_\perp )
={2(\gamma -1)\over (\gamma +1)^2}\,{\mu \over k}\,m_p V^2_\perp
\simeq 3\times 10^7\left({V_\perp \over 10^3\,\,{\rm km\,s}^{-1}}
\right)^2\,\,\,{\rm K}\,, \eqno(1)$$
where $\gamma$ is the ratio of heat capacities at constant pressure
and  at constant volume, $\mu$=$A/(1+Z)$ is the mean molecular
weight, $A$ is the atomic
weight of ion, $Z$ is its electrical charge, $k$ is the Boltzmann
constant, $m_p$ is the proton mass, and $V_\perp$ is the component
of the stellar-wind velocity perpendicular to the shock front. The
value $\gamma$ is equal to 5/3 for a rarefied  totally-ionized plasma.
If helium predominates in the gas of the WR stellar wind, we have
$A=4; Z=2; \mu = 4/3$.
\par Behind the shock the hot gas outflows from the region of
stellar wind
collision nearly along the contact surface (see in detail Usov 1992).
\par In the case of the collision of two spherical winds with the
terminal velocities the distances
${r_{_{\rm WR}}}$ and ${r_{_{\rm
OB}}}$ from WR or OB stars, respectively, to the region
where these winds meet is
$$r_{_{WR}}={1\over 1+\eta^{1/2}}D,\qquad r_{_{OB}}={\eta^{1/2}\over
1+\eta^{1/2}}D\eqno(2)$$
(see Fig. 1), where
$$\eta={\dot M_{_{OB}}V^\infty_{_{OB}}\over\dot M_{_{WR}}V^
\infty_{_{WR}}}\, .
$$
\par Since $\dot M_{_{OB}}\sim (0.01-0.1)\dot M_{_{WR}}$ and
 $V^\infty_{_{OB}}\sim
V^\infty_{_{WR}}$, the dimensionless parameter $\eta$ is small
 ($\eta\ll 1$). Hence,
the region of stellar wind collision is much nearer to the OB star than
 to the WR
star ($r_{_{WR}}\gg r_{_{OB}}$).
\par The form of the contact surface $C$ near the OB star $( r \sim
r_{_{OB}} )$ is given by (see, e.g., Usov 1992)
$$\vert \vec R_c(\chi )\vert
\simeq r_{_{OB}} {\chi \over \sin \chi} \eqno (3)$$
for $\chi \,\leq\, \pi/2 $ ( here $\chi$ is the angle between the radius
vector $\vec R_c(\chi )$
from the center of the OB star to the point at the contact surface
and the line connecting the components of the binary).
\par At intermediate
 distances $( r_{_{OB}} \ll r < (P/2) V^\infty _{_{WR}})$
from the OB star the contact surface $C$ approaches the conic surface
$\tilde C$  with angle
$$\theta\simeq 2.1\,\left
(1-{\eta^{2/5}\over 4}\right) \eta^{1/3}\,\,\,\,\,\,\,\,{\rm for}\,\,
10^{-4}\leq\eta\leq 1\,. \eqno(4)$$
\par The shock fronts $S_1$ and $S_2$ are near the conic surfaces $\tilde
S_1$ and $\tilde S_2$
at $r_{_{OB}} \ll r< (P/2) V^\infty _{_{WR}}$
as well. In the case of wide WR+OB binaries, like WR 140, when the
 fraction of the
thermal energy of the hot gas lost by emission in the shock layers
 is small the angle
$\Delta\theta$ between these conic surfaces is $\sim\theta$.
In the other case the value of $\Delta \theta$
is smaller than $\theta$ and asymptotically approaches zero
if this fraction goes to unity.
\par The contact surface $C$ and two shock fronts $S_1$ and $S_2$
are twisted at a distance more than  $\sim (P/2)V^\infty_{_{WR}}$
due to the orbital rotation and have spiral form (see Fig.
1). Since the orbital velocities in massive
binaries are typically much smaller than the terminal velocities we have
$(P/2)V^\infty_{_{WR}}\gg D$.
\smallskip
\par Since both $r_{_{WR}}\gg r_{_{OB}}$ and the
radius of OB star $R_{_{OB}}$ is , as a rule, essentially  larger
than the radius of the WR star, in close binaries the gas velocity of OB
star wind does not reach the terminal value before the shock $S_2$.
\par The velocity of the matter outflow $V_{_{OB}}(r)$ at  distance $r$
from the center of the OB star varies from almost zero on the OB
star surface to $V^{^\infty}_{_{OB}}$ for $r>r_{ter}$ (Barlow 1982),
where $r_{ter}$ is approximately equal to (3-5) $R_{_{OB}}$.
\par If $r_{_{OB}}>r_{ter}$, we have the collision of two winds with
terminal velocities and values $r_{_{WR}}$ and $r_{_{OB}}$ are determined
by equation (2).  If $R_{_{OB}}<
r_{_{OB}}<r_{ter}$, the gas flowing out of the OB
star does not reach the terminal velocity by the time it enters the
shock.  In turn, the stellar-wind gas of the WR star is decelerated by
the radiation pressure of the OB star while approaching
the OB star. In this case, the gas velocities ahead of and near the
shocks drop, and the gas temperature behind the shocks drops too
(Usov 1990, 1992).
If $r_{_{OB}}\leq R_{_{OB}}$, the OB wind is suppressed by the
ram pressure of the WR wind, on the side facing the WR stars.

\medskip
{2.3. {\it Basic equations and boundary conditions}}
\smallskip
\par The set of equations which describes the steady gas flow between the
shock wave and the contact surface will be the continuity equation:
$$ {\rm div}\,(\rho\vec V)=0\, ,\eqno(5)$$
the momentum equation:
$$\rho(\vec V\nabla)\vec V = - \nabla p \ , \ \eqno(6)$$
and the energy equation:
$$\rho\vec V\nabla (H + |\vec V|^2/2) = - Q \, \,, \eqno(7)$$
where $\rho$ is the density of the gas and $Q$ is the energy loss
per unit gas volume by radiation.
\par Since the gas in the shock layer is practically totally ionized, its
pressure $p$ and its specific enthalpy $H$ can be expressed as
$$p = (N_e + N_i)kT= {\rho kT\over m_p\mu} \ , \ \eqno(8)$$
$$H = {\gamma\over \gamma - 1}{p\over\rho} = {5\over 2}{kT\over m_p\mu} \
, \ \eqno(9)$$
here $N_i = \rho/m_pA$ is the ion density, $N_e = N_iZ$ is the electron
density.
\par The ionized gas heated in the
shock layer is emitting mainly due to free-free transitions of electrons
in the Coulomb fields of ions at the gas temperature $T >
3 \times 10^6$ K (Gaetz and Salpeter 1983 ).
 \par Gas parameters ahead of the shock front (index 1) and behind the
shoch front (index 2) are related via the Rankin-Hugoniot relations
$$\rho_1V^{(n)}_1=\rho_2V^{(n)}_2,
\,\,\,\,\,\, p_1+\rho_1 (V^{(n)}_1)^2=p_2+\rho_2
(V_2^{(n)} )^2 \ , $$
$$V_1^{(\tau)}=V_2^{(\tau)},\, \,\,\,\, \, \, H_1+{1\over
2}(V_1^{(n)})^2=H_2+{1\over 2}(V_2^{(n)})^2\,. \eqno(10)$$
Indices $n$ and $\tau$ denote the normal and tangential components of the
vector $\vec V$.  The condition $V^{(n)}$=0 is met on the contact
surface.  This condition and the Rankine-Hugoniot relations (10) are the
full set of boundary conditions for the set of equation (5)-(9) needed
to find the parameters of hot gas in the shock layers.
\medskip
2.4. {\it The method of solution}
\smallskip
\par The set of equations (5)-(9) with the boundary conditions
(10) was solved in (Galeev, Pilyugin and Usov 1989; Bairamov,
Pilyugin and Usov 1990; Usov 1992)
analytically using the method of Chernyi (1961). In this method
the ratio of the density of the gas, $\epsilon$, ahead of the shock
to that behind it was considered as a small parameter, $\epsilon
\ll 1$. If the fraction of the thermal energy of the hot gas lost by
emission in the shock layer is small, $\epsilon$ is equal to
$(\gamma -1)/(\gamma +1)=1/4$. If the thermal energy losses are
essential, the value of $\epsilon$ decreases and goes to zero when
practically all thermal energy of the hot gas is radiated.
\par The problem of stellar wind collision was solved in the following
way. First, the parameters of the hot gas in the external and internal
shock layers were obtained as a function of the form of unknown contact
surface. Then the form of the contact surface was found
from the equality of the gas pressure on both sides of contact surface.
\par In the Newtonian approximation, when we neglect the Busemann
correction, the equation for the contact surface is given by
equation (3). The difference between the analytical equation (3) and
the numerical solution for the contact surface with Busemann correction
is smaller than 10\%. Therefore, to calculate both the parameters of
the hot gas in the external and internal shock layers and X-ray
emission from the shock layers equation (3) for the contact surface
was used (Usov 1992).

\heading{3. X-RAYS FROM THE REGION OF STELLAR WIND COLLISION}
\medskip
3.1. {\it Collision of two stellar winds with terminal velocities}
\smallskip
\par If a WR + OB binary is wide enough, $r_{_{OB}}>r_{ter}$,
the WR and OB winds collide with the terminal velocities, and the
fraction of the thermal energy of the hot gas lost by emission in
the shock layers is, as a rule, small. In this case, the expected
X-ray luminosity is (Usov 1992)
$$L_{ext}\simeq 8\times 10^{34}\left({\dot M_{_{WR}}\over
10^{-5}M_\odot yr^{-1}}\right)^{1/2}\left({\dot M_{_{OB}}\over
10^{-6}M_\odot yr^{-1}}\right)^{3/2}
\left({V^{^\infty}_{_{WR}}\over 10^3\,km\,s^{-1}}\right)^{-{5/2}}$$
$$\times \left({ V^{^\infty}_{_{OB}}\over 10^3\,km\,s^{-1}}\right)^{3/
2}\left({D\over 10^{13}cm}\right)^{-1}\,\,\,{\rm erg\,\,s}^{-1}
\eqno(11)$$
from the external shock layer between the shock front $S_1$ and
the contact surface (see Fig. 1), and
$$L_{int}\simeq 1.3\times 10^{35}\left({\dot M_{_{WR}}\over
10^{-5}M_\odot yr^{-1}}\right)^{1/2}\left({\dot M_{_{OB}}\over
10^{-6}M_\odot yr^{-1}}\right)^{3/2}\left({V^{^\infty}_{_{WR}}\over
10^3km\,s^{-1}}\right)^{1/2}$$
$$\times\left({V^{^\infty}_{_{OB}}\over 10^3km\,s^{-1}}\right)^{-{3/2}}
\left({D\over 10^{13}cm}\right)^{-1}\,\,\,{\rm erg\,\,s}^{-1}
\eqno(12)$$
from the internal shock layer between the shock front $S_2$ and the
contact surface.
\par The total power of X-ray emission from the region of stellar wind
collision is $L_x = L_{ext}+L_{int}$.
The X-ray spectra of the shock layers may be roughly approximated by
the bremsstrahlung radiation of the uniform totally ionized plasma
with the temperature $\sim (0.8- 0.9)T(V^{^\infty}_{_{WR}})$
for the external shock layer and $\sim (0.8-0.9)T(V^{^\infty}_{_{OB}})$
for the internal shock layer, where $T$ is given by equation (1).
\par To calculate the parameters of hot gas in the shock layers and the
X-ray emission of this gas the values of $\epsilon$ and $\eta^{1/2}$
were considered as small parameters. Since the zero and first terms
of the series were taken into account only, the accuracy of our
calculations is of the order of $(\epsilon^2+\eta)^{1/2}$, i.e. the
accuracy is several tens of percent for $\eta\leq  0.1$ which is
typical  for WR+OB binaries.
\par In our consideration of X-ray emission from the region of stellar
wind collision we did not take into account the absorption of X-rays in
the stellar winds. We can do it for wide WR + OB binaries,
$D> 10^{13}$ cm, (Usov 1992) when the sightline to the region of stellar
wind collision is passed far enough from the WR and OB stars.
\medskip
3.2. {\it Close and very wide binaries}
\smallskip
\par In a close binary, $r_{_{OB}}<r_{ter}$, both the WR stellar wind
and the OB stellar wind have velocities near the shock wave which are
less than their terminal velocities (see Section 2.2). To estimate
the X-ray emission from the region of stellar wind collision
it is necessary to substitute
$V^{^\infty}_{_{WR}}$ and $V^{^\infty}_{_{OB}}$ for $V^*_{_{WR}}$ and
$V^*_{_{OB}}$ in equations (11) and (12) (here $V^*_{_{WR}}$ and
$V^*_{_{OB}}$ are the gas velocities of the WR and OB stellar winds,
respectively, ahead of and near the shocks). Let us discuss briefly
the behaviour of the X-ray luminosity, $L_x$, for a remote observer
when the value of $r_{_{OB}}$ varies from $r_{ter}$ to $R_{_{OB}}$.
\par At first the X-ray luminosity $L_x$ may increase a few times in
comparison with the sum of equations (11) and (12),
$L_x\propto (V^*_{_{WR,OB}})^{-1}$. In turn, the gas temperature
in the shock layers drops, $T\propto (V^*_{_{WR,OB}})^2$, and the
spectrum of the X-ray emission shifts to the soft region (Usov 1990).
Then, if the value of $r_{_{OB}}$ is a few times smaller
than $r_{ter}$, the main portion of this soft X-ray emission is
absorbed in the WR and OB stellar winds and the X-ray luminosity is
decreased (Luo, McCray and Low 1990; Usov 1990). At $1\leq (r_{_{OB}}
/R_{_{OB}})\leq 1.2-1.3$ the X-ray emission is
suppressed practically at all. Hence, it is expected
that the X-ray luminosity of a WR + OB binary is as high as possible
when $r_{_{OB}} \simeq (0.5-1)r_{ter}$ or $D \simeq
(0.5-1) {\eta}^{-1/2}r_{ter}$.
\par It is worth noting that some increase of the X-ray flux may be
observed near the conic surface $\tilde C$ from more or less close
WR + OB binaries for which the X-ray absorption in the stellar
winds is essential. The point is that the outflowing plasma in the
shock layers is highly ionized, and the X-ray absorption along
this surface may be small.
\par Equations (11) and (12) are valid only if there is temperature
equalization between ions and electrons in the shock layers.
If there is no such an equalization, the power of the
bremsstrahlung emission from the layer is
$\sim L(D/D_{eq})^{-1/3}$ at $D\geq D_{eq}$, where $L$
is given by either equation (11) or equation (12) depending on the kind
of the shock layers and $D_{eq}$ is the distance between the conponents
of the binary at which temperature equalization in the shock layer
is broken (Usov 1992). The temperature of electrons which determines
the X-ray spectrum is $\propto(D/D_{eq})^{-2/3}$ at $D\geq D_{eq}$.
\smallskip
\par Among WR + OB binaries, WR 140 is the most powerful X-ray
source, $L_x\simeq 4\times 10^{34}$ ergs s$^{-1}$ (Pollock 1987).
It is natural because at the moment of the X-ray observation
WR 140 was more or less near the state with $r_{_{OB}}\sim r_{ter}$.
The observed value of $L_x$ is in agreement with the X-ray luminosity
which is expected from the region of stellar wind collision in WR 140
(Usov 1992).
\smallskip
\par Except for the X-ray emission generated in the region of stellar
wind collision, WR + OB binaries may have an X-ray emission which
is inherent in single stars.  X-ray emission of single
massive stars may be generated due to heating
of the outflowing gas to temperatures of about a few million degrees
either by a large number of quasi-periodic strong shocks (Lucy and
White 1980; Cassinelli and Swank 1983; Owocki and Rybicki 1985) or by
the current sheets (Usov and Melrose 1992). From analysis of time
variability of X-ray emission from WR + OB binaries it is
possible to separate X-ray emission of colliding winds
from X-ray emission that is inherent in single stars.

\heading{4. GAS COOLING AND DUST FORMATION}
\par In the shocks the gas of stellar winds is compressed but not more
than $\sim$ 4 times.  If a WR + OB binary is close enough and
during the gas motion through the shock layer
the main part of the thermal energy of the hot gas is radiated, the
degree of the gas compression can increase significantly,
i.e. a cold dense gas may be formed.
The terminal velocity of the cold gas has to be essentially smaller
than $V^{^\infty}_{_{WR}}$ and $V^{^\infty}_{_{OB}}$. This cold gas
may be heated and accelerated because of its interaction with the WR
high-velocity wind.
\par In the
case of long-period binaries like WR 140 even near the periastron passage
the fraction of the thermal energy of the hot gas lost by emission is
small.  At first sight in this case the hot gas outflowing in the shock
layer has to expand quasi-adiabatically, and the essential cooling and
compression of the gas in the region of
the stellar wind collision is impossible.
But it is not so.  Indeed the main part of the hot gas in the shock layer
expands quasi-adiabatically.  However, near the contact surface there
is the region where the energy losses may be important and the gas may
cool and compress very strongly (Usov 1991).
The strong cooling in this region is caused by the slow motion of the gas
along the contact surface, such that the gas has ample time to cool.
In this case the fraction of the WR wind which may be strongly cooled
and compressed in the shock layers is (Usov 1991)
$$\alpha\simeq \eta ^2\left({\dot M_{_{WR}}\over 10^{-5}\,M_\odot
yr^{-1}}\right)^2\left({D\over 10^{13}\,cm}\right)^{-2}\left(
{V^*_{_{WR}}\over 10^8\,cm\,s^{-1}}\right)^{-6}\,.\eqno(13)$$
\par The characteristic cooling time of the hot plasma via the
bremsstrahlung depends on the temperature and the density as
$\tau_c\simeq 1.5\times 10^{11}N^{-1}T^{1/2}$ s. Taking into account
that the external pressure of the hot gas compresses the cooling gas
($N\propto T^{-1}$), we have $\tau _c \propto T^{3/2}$, i. e.
the process of cooling near the contact
surface is accelerating (thermal instability). The gas near the contact
surface may be cooled down to $\sim 10^4 - 10^5$ K, and the density
of the cold gas may be as large as $\sim 10^3-10^4 \rho _*$, where
$\rho _*$ is the gas density before the shock (Usov 1991).
\par At the distance $r_d\simeq$ a few $\times 10^{15}$ cm from the
binary, the cold dense gas can form a dust shell and reradiate
the UV radiation in the IR region (Williams et al. 1990, 1992, 1994;
Usov 1991).
\par The value of $\alpha$ is, as a rule, very small at $r_{_{OB}} >
r_{ter}$ because of high velosity of the stellar winds and increases
very sharply at $r_{_{OB}} < r_{ter}$. When $r_{_{OB}}$ is
comparable with $R_{_{OB}}$, the dynamical gas pressures of both
winds ahead of the shocks are decreased, and the gas compression in
the shock layers is decreased too. In addition, in close WR + OB
binaries with $r_{_{OB}}\sim R_{_{OB}}$ the X-ray absorption inside
the shock layers is high, and it can prevent a strong gas cooling in
the layers. Therefore, when the value of $r_{_{OB}}$ is somewhere
between $\sim r_{ter}$ and $\sim 0.3 r_{ter}$ (say, $r_{_{OB}}\simeq
0.5 r_{ter}$), the most favourable conditions for a strong compression
of the gas in the shock layers are realized (Usov 1991).
\par  The long-period WR binaries (WR 48a, WR 125, WR 137, WR 140, and
possibly WR 19) with a large enough value of the orbital eccentricity
are distinguished among the rest of the WR binaries in that they
experience the state with $r_{_{OB}}\sim 0.5 r_{ter}$ in the process
of their orbital motion. In this state there is a strong outflow of
cold gas, which leads to an IR outburst (Williams et al. 1992, 1994).

\heading{5. PARTICLE ACCELERATION AND RADIO AND $\gamma$-RAY EMISSION}
\par It is well known that plasma shocks in astrophysical setting can
and do accelerate charged particles to high energies (for a review, see
Blandford and Eichler 1987  and Ellison and Jones 1991 and references
therein). Therefore the strong shocks formed by the
colliding stellar winds in WR + OB binaries are very promising as
particle accelerators.
\par Necessary conditions for particle  acceleration by
a shock wave include the following (Eichler and Usov 1993):
\smallskip
\par 1. {\it The shock must be collisionless in the
absence of an external injection mechanism.}
This is typically satisfied if the strength of the
magnetic field ahead of the shock is $B\gg 10^{-6}$ Gauss. Here
and below all numerical estimates are given for a typical WR + OB binary
(see Section 2.1).
\par 2. {\it The Ohmic damping rate of Alfven waves must not exceed the
maximum growth rate}.
\par 3. {\it For primary electron acceleration, the shock velocity must
considerably exceed the phase velocity of whistler waves propagating
normal to the shock.}
\smallskip
\par Last two conditions holds if $B\ll$ a few Gauss.
\par All of these constraints on B are typically  satisfied by colliding
winds in WR + OB binaries. Hence the region of stellar wind collision
in these binaries may be a strong source both high energy particles
and nonthermal radiation generated by these particles.
\par  The maximum electron  energy is limited by
ion-neutral wave damping and by inverse Compton losses (Bell 1978).
The maximum energy allowed by ion neutral damping should be well
above 100 Gev and does not play an important role.
\par The upper limit on the Lorentz factor of electrons
due to inverse Compton losses can be shown, assuming a quasi-parallel
shock geometry, to be given by (Eichler and Usov 1993)
$$\Gamma_{\rm max}^2 \leq { 3\pi eBcr^2_{_{OB}}\over \lambda \sigma _
{_{T}}L_{bol}} \left ({V^\infty_{_{WR}} \over c} \right )^2 \simeq
3 \times 10^8 \eta
\left ({B \over {\rm Gauss}}\right )\,,
\eqno(14)$$
with approximate equality obtaining in the absence of any other,
stronger limits. Here $L_{bol} \simeq 10^{39}$ ergs s$^{-1}$
is the bolometric luminosity of the $OB$ star,
$\sigma _{_{T}}$ is the cross-section of Thomson scattering, $e$ is
the electron charge and $\lambda \simeq 3$ is the ratio of mean free
path to gyroradius.
\par Given the constraints of $B$ discussed above, the energy of
electrons $E_{\rm max}=\Gamma_{\rm max}mc^2$ is enough to
generate synchrotron radio emission at the frequencies of which
nonthermal radio emission from WR + OB binaries has been observed
(Florkowski and Gottesman 1977; Becker and White 1985;
Abbott et al. 1986; Felli and Massi 1991).
\par A differential energy spectrum expected for relativistic
electrons accelerated in the region of stellar wind collision is
$N(E) \propto E^{-\beta}\,$,
where $N(E)dE$ is the number of high energy electrons per unit
volume with total energy $E$ in the interval $dE$ and $\beta \simeq 2$.
The spectrum of synchrotron radiation generated by relativistic
electrons with this energy spectrum  is
$S(\nu) \propto \nu ^{-\gamma}\,,$
here $S(\nu )$ is the flux of radiation at the frequency $\nu $ and
$\gamma = (\beta  -1)/2 \simeq 0.5$ is a spectral index.
\par
Typically, only a small  fraction of the energy of accelerated electrons
are converted to radio emission before the flow convects them out of
the colliding wind region.
\par
{}From equation (3) we can get the characteristic size of nonthermal radio
emisson
$$l^{rad} \simeq 2 R_c(\pi /2) \simeq \pi r_{_{OB}} \,. \eqno (15)$$
\par Stellar winds outflowing from the WR and OB stars are opaque to
free-free absorption. A typical radius of the radio
photosphere at the frequency of a few GHZ for WR stars
is $R^{rad}_{_{WR}}\simeq$ a few times $10^{14}$cm (Wright and
Barlow 1975). The value of this radius for OB stars, $R^{rad}_{_{OB}}$,
is approximately an order of magnitude smaller than $R^{rad}_{_{WR}}$.
Hence, in the case of WR and OB stars the radius of their radio
photosphere is of the order of hundred  or more stellar radii.
\par Radio emission from the colliding wind region
may be observed at the frequency $\nu$ only if the time
$R^{rad}_{_{WR}}/V^\infty_{_{WR}}$ during which the gas outflowing from
the WR star with the velocity $V^\infty_{_{WR}}$ reaches the radius of
the radio photosphere $R^{rad}_{_{WR}}$ is of the order of or smaller
than the time which is necessary for the WR+OB binary to turn on the
angle $2\theta+\Delta\theta\simeq 3\theta$ (see Fig. 1), i.e.
$${R^{rad}_{_{WR}}\over V^\infty_{_{WR}}} \leq
{3\theta\over 2\pi} P\,\,\,\,\,\,{\rm or}\,\,\,\,\,\,P\geq
P_{cr}={2\pi\over 3\theta}{R^{rad}_{_{WR}}\over V^
\infty_{_{WR}}}\simeq 13\eta^{-{1/3}}
\left({\nu\over 5\, {\rm GHz}}\right)^{-{2/3}}\; {\rm
days}\,.\eqno(16)$$
\par For a typical WR+OB binary (see Section 2.1) the value of $P_{cr}$
is of the order of a month  when the frequency of observation is a few
GHz.
\par If $P<P_{cr}$ nonthermal radio emission from the region of stellar
wind collision in the WR+OB binary cannot be observed because it is
screened by the WR wind.
\smallskip

\par
Striking variations of radio emission from remarkable WR binary
WR 140 were observed (Florkowski and Gottesman 1977;
Becker and White 1985; Abbott et al. 1986). This time variability
and the other observational data on radio emission of WR 140,
such as luminosity, spectrum,
the size of nonthermal radio source, etc., may be
explained by a model in which its nonthermal radio emission is
generated at the site of stellar wind collision (Eichler and Usov 1993).
\par Gamma-ray emission can be attributed to either inverse
Compton emission or $p-p$ collisions followed by pion decay.
In the former case, the seed photons would presumably be the
processed UV radiation from the OB stars. A high-energy tail of the
inverse Compton spectrum of $\gamma$-rays may exist
up to $\sim 10^3$ MeV (Eichler and Usov 1993).
\par Considerations of shock acceleration theory suggest
that the  WR 140 system may be a strong source of $\gamma$-rays
(Eichler and Usov 1993). The $\gamma$-ray flux has to increase
essentially near the periastron passage. Recently, the hard X-ray and
soft $\gamma$-ray emission from WR 140
was observed by the OSSE detector aboard the CGRO (Hermsen et al. 1994).

\heading{REFERENCES}

\noindent
Abbott, D.C., Bieging, J.H., Churchwell, E., and Torres, A.V. 1986, Ap.
J., {\bf 303}, 239.

\noindent
Abbott, D.C., and Conti, P.S. 1987, Ann. Rev. Astr. Ap.,
{\bf 25}, 113.

\noindent
Bairamov, Z.T., Pilyugin, N.N., and Usov, V.V. 1990, Soviet Astron.,
{\bf 34}, 502.

\noindent
Barlow, M.J. 1982, in IAU Symposium 99, Wolf-Rayet stars: observations,
physics, evolution, ed. C.W.H. de Loore and A.J. Willis (Dordrecht:
Reidel), p. 149.

\noindent
Becker, R.H., and White, R.L. 1985, Ap. J., {\bf 297}, 649.

\noindent
Bell, A.R. 1978,  M.N.R.A.S, {\bf 182,} 147.

\noindent
Bianchi, L. 1982, Astr. Sp. Sci., {\bf 82}, 161.

\noindent
Blandford, R.D., and Eichler, D. 1987, Physics Reports, {\bf 154}, 1.

\noindent
Caillaut, J.P., Chanan, G.A., Helfand, D.J., Patterson, J., Nousek, J.A.
Talacko, L.O., Bothun, G.D., and Becker, R.H. 1985, Nature, {\bf 313},
376.

\noindent
Cassinelli, J.P., and Swank, J.H. 1983, Ap. J., {\bf 271}, 681.

\noindent
Cherepashchuk, A.M. 1976, Soviet Astr. Lett., {\bf 2}, 138.

\noindent
Chernyi, G.G. 1961, Introduction to Hypersonic Flow (New York, London:
Academic Press).

\noindent
Cooke, B.A., Fabian, A.C., and Pringle, J.E. 1978, Nature, {\bf 273},
645.

\noindent
Eichler, D., and Usov, V.V. 1993, Ap. J., {\bf 402}, 271.

\noindent
Ellison, D.C., and Jones, F.C. 1991, Sp. Sci. Rev., {\bf 58}, 259.

\noindent
Felli, M., and Massi, M. 1991, Astr. Ap., {\bf 246}, 503.

\noindent
Florkowski, D.R., and Gottesman, S.T. 1977, M.N.R.A.S.,
{\bf 179}, 105.

\noindent
Gaetz, T.J., and Salpeter, E.E. 1983, Ap. J. Suppl., {\bf 52}, 155.

\noindent
Galeev, A.A., Pilyugin, N.N., and Usov, V.V. 1989, in Proc.
Varenna - Abastumani International
  School and Workshop on Plasma Astrophysics,
Varenna, Italy, p. 125.

\noindent
Garmany, C.D., and Conti, P.S. 1984, Ap.  J., {\bf 284}, 705.

\noindent
Hermsen, W. et al. 1994, in Proc. Second Compton Symposium (in press).

\noindent
Leitherer, C. 1988, Ap. J., {\bf 326}, 356.

\noindent
Lucy, L.B., and White, R.L. 1980, Ap. J., {\bf 241}, 300.

\noindent
Luo, D., McCray, R., and Mac Low, M.-M. 1990, Ap. J. {\bf 362}, 267.

\noindent
Moffat, A.F.J., Firmani, C., McLean, I.S., and Seggewiss, W. 1982, in IAU
Symposium 99, Wolf-Rayet stars: observations, physics, evolution, eds.
C.W.H. de Loore and A.J.  Willis (Dordrecht: Reidel), p. 577.

\noindent
Myasnikov, A.V., Zhekov, S.A. 1993, MNRAS, {\bf 260}, 221.

\noindent
Owocki, S.P., and Rybicki, G.B. 1985, Ap. J., {\bf 299}, 265.

\noindent
Pollock, A.M.T. 1987, Ap. J., {\bf 320}, 283.

\noindent
Prilutskii, O.F., and Usov, V.V. 1975, Astron. Cirk, No 854, 1.

\noindent
Prilutskii, O.F., and Usov, V.V. 1976, Soviet Astron., {\bf 20}, 2.

\noindent
Seward, F.D., Forman, W., Giacconi, R., Griffiths, R., Harnden, F.R.,
Jr., Jones, C., and Pye, J. 1979, Ap. J. (Letters), {\bf 234}, L55.

\noindent
Stevens, I.R., Blondin, J.M., and Pollock, A.M.T. 1992, Ap. J.
{\bf 386}, 265.

\noindent
Torres, A.V., Conti, P.S., and Massey, P. 1986, Ap. J.,
{\bf 300}, 379.

\noindent
Usov, V.V. 1990, Astr.  Sp. Sci., {\bf 167}, 297.

\noindent
Usov, V.V. 1991, M.N.R.A.S., {\bf 252}, 49.

\noindent
Usov, V.V. 1992, Ap. J., {\bf 389}, 635.

\noindent
Usov, V.V., and Melrose, D.B. 1992, Ap. J., {\bf 395}, 575.

\noindent
Williams, P.M. et al. 1992, MNRAS, {\bf 258}, 461.

\noindent
Williams, P.M., van der Hucht, K.A., Kidger, M.R., Geballe, T.R.,
and Bouchet, P. 1994, MNRAS, {\bf 266}, 247.

\noindent
Williams, P.M., van der Hucht, K.A., Pollock, A.M.T., Florkowski, D.R.,
van der Woerd, H., and Wamsteker, W.M. 1990, M.N.R.A.S.,
{\bf 243}, 662.

\noindent
Williams, P.M., van der Hucht, K.A., and Th\'e, P.S.
1987, Astr. Ap., {\bf 182}, 91.

\noindent
Willis, A.J. 1982, M.N.R.A.S., {\bf 198}, 897.

\noindent
Wright, A.E., and Barlow, M.J. 1975, M.N.R.A.S., {\bf 170}, 41.

\end